\documentclass[fleqn,usenatbib]{mnras}

\usepackage{newtxtext,newtxmath}
\usepackage{ulem}

\usepackage[T1]{fontenc}
\usepackage{ae,aecompl}

\usepackage{graphicx}	
\usepackage{amsmath}	
\usepackage{amssymb}	
\hypersetup{draft}     

\newcommand{\Msun}{$\mathrm{M}_\odot$}
\newcommand{\Rsun}{$\mathrm{R}_\odot$}

\title[PISN models for OGLE14-073]
{OGLE14-073 -- a promising pair-instability supernova candidate}

\author[A. Kozyreva et al.]{Alexandra~Kozyreva\,$^{1}$\thanks{E-mail: sasha@wise.tau.ac.il},
Markus Kromer\,$^{\,2,3}$, Ulrich~M.~Noebauer\,$^{\,4}$, 
\newauthor
Raphael~Hirschi\,$^{\,5,6}$
\\
$^{1}$The Sackler School of Physics and Astronomy, Tel Aviv University, 69938 Tel Aviv, Israel\\
$^{2}$Zentrum f\"ur Astronomie der Universit\"at Heidelberg, Institut f\"ur Theoretische Astrophysik, D-69120 Heidelberg, Germany\\
$^{3}$Heidelberger Institut f\"ur Theoretische Studien, D-69118 Heidelberg, Germany\\
$^{4}$Max-Planck-Institut f\"ur Astrophysik, Karl-Schwarzschild-Stra{\ss}e 1, D-85748 Garching, Germany\\
$^{5}$Astrophysics group, Keele University, Keele, Staffordshire, ST5 5BG, UK\\
$^{6}$Kavli IPMU (WPI), University of Tokyo, Kashiwa, Chiba 277-8583, Japan\\
}

\date{Accepted XXX. Received YYY; in original form ZZZ}

\pubyear{2018}

\begin{document}
\label{firstpage}
\pagerange{\pageref{firstpage}--\pageref{lastpage}}
\maketitle

\begin{abstract}
  The recently discovered bright type\,II supernova OGLE14-073
  evolved very slowly.  The light curve rose to maximum for 90~days
  from discovery and then declined at a rate compatible with the
  radioactive decay of $^{56}$Co.  In this study, we show that a
  pair-instability supernova is a plausible mechanism for this
  event. We calculate explosion models and light curves with the
  radiation hydrodynamics code {\sc stella} starting from two
  M$_{\rm ZAMS}$ = 150\,\Msun{}, \textit{Z}=0.001 progenitors. We
  obtain satisfactory fits to OGLE14-073 broadband light curves by
  including additional $^{56}$Ni in the centre of the models and
  mixing hydrogen down into the inner layers of the ejecta to a 
  radial mass coordinate of 10\,\Msun{}.  The extra
  $^{56}$Ni required points to a slightly more massive progenitor
  star. The mixing of hydrogen could be due to large scale mixing
  during the explosion.  We also present synthetic spectra for our
  models simulated with the Monte Carlo radiative transfer code {\sc
    artis}.  The synthetic spectra reproduce the main features of the
  observed spectra of OGLE14-073. We conclude that OGLE14-073 is one
  of the most promising candidates for a pair-instability explosion.
\end{abstract}




\section[Introduction]{Introduction}
\label{sect:intro}

OGLE14-073 is a unique supernova (SN) event which was discovered long
before its peak \citep{2014ATel.6489....1B, 2014ATel.6494....1W,
2017NatAs...1..713T}.  It was classified as a bright hydrogen-rich
type II supernova at redshift z=0.1225. Thanks to an extensive
follow-up campaign, it has good data coverage, starting from 90~days
before peak magnitude in the \textit{I}-band. Recently,
\citet{2018MNRAS.475L..11M} and \citet{2017arXiv171204492D}
suggested fallback accretion or a magnetar
as the powering engine of this bright SN.

However, the long rise to peak could also indicate a high ejecta mass
(\citealt{2017NatAs...1..713T} estimate
M$_\mathrm{ej}\sim 60$~\Msun{}). This can only be realized if the
initial mass of the progenitor star was above at least
100~\Msun{}. From a theoretical point of view, it is well known that
stars with initial mass range between 140\,--\,260~\Msun{} undergo
pair instability and explode as pair-instability supernovae
\citep[hereafter PISN,][]{1967PhRvL..18..379B,1967ApJ...148..803R}.
Depending on the amount of radioactive $^{56}$Ni produced, the
supernova may be either sub- or superluminous
\citep{2017ApJ...846..100G}.

The long rise of OGLE14-073 and the presence of H Balmer
lines throughout observations covering more than 150~days
require a massive H-rich ejecta. 
The peak luminosity in combination with an assumption about
the time of explosion constrains the amount of $^{56}$Ni which powers
the light curve \citep{1982ApJ...253..785A}. Unfortunately, the
explosion time is quite uncertain for OGLE14-073. According to the
findings by \citet{2016MNRAS.459L..21K}, the peak luminosity of
$9.6\,\pm\,0.2\times\,10^{\,42}$\,erg\,s$^{\,-1}$ for OGLE14-073
\citep{2017NatAs...1..713T} requires 1.3~\Msun{} of
$^{56}$Ni. \citet{2017NatAs...1..713T} derive a lower limit for the
$^{56}$Ni mass of 0.47~\Msun{}$\pm0.02$~\Msun{} assuming that the
explosion occurred just before the first detection.

In a pair-instability explosion, there is a strong correlation between
the mass of the final carbon-oxygen-rich core (CO core) and the $^{56}$Ni
yield \citep[see e.\,g.][and references therein]{2002ApJ...567..532H},
because nickel production depends on the degree of compression during
the collapse phase.  The larger the CO core mass, the deeper the
gravitational potential reached during the pair-instability phase. In
turn, higher density and temperature are reached at maximum
compression, releasing more nuclear energy and producing higher yields
of $^{56}$Ni. Most of the helium-rich core is burnt into
carbon and oxygen in PISN progenitors, leaving a narrow He-shell on
top of the core, which only amounts to a few solar masses \citep[see
e.\,g.][]{2013MNRAS.433.1114Y}. Thus the final helium core mass
roughly equals the CO core mass. In PISN models, stars with helium cores of
80\,--\,90~\Msun{} generate 0.5-1.3~\Msun{} of $^{56}$Ni. Note
that the nickel yield varies rapidly at the lower end of the PISN mass range.
For example, a model with a 70~\Msun{} He core (CO-core of 64~\Msun{}) gives
0.02~\Msun{} of $^{56}$Ni, while a model with a He core of 90~\Msun{} (CO-core
of 82~\Msun{}) produces already 1.3~\Msun{} of
$^{56}$Ni. This implies that a 25\% increase in the helium core mass
corresponds to almost two orders of magnitude difference in the $^{56}$Ni yield
\citep{2002ApJ...567..532H}. Hence, assuming that OGLE14-073 exploded as a
PISN, its progenitor is expected to be a very massive star with a final helium
core mass in the range of 80\,--\,90~\Msun{}.

In the present study, we examine a possible PISN origin of
OLGE14-073. In particular, we construct ejecta configurations based on
two existing self-consistent PISN models to explain the light curves
and spectra of OGLE14-073. Details of our explosion models and the 
techniques to model light curves and spectra are given in 
Section~\ref{sect:method}. In Section~\ref{sect:results}, we present 
synthetic observables for our best-fitting model and discuss how the light
curve properties depend on different progenitor and explosion
parameters before concluding the study in
Section~\ref{sect:conclusion}.

\section[Progenitor models and modeling of light curves]
{Modelling OGLE14-073}
\label{sect:method}

\subsection[The choice for the evolutionary models]
{Explosion models}
\label{subsect:method1}

\begin{table*}
\centering
\caption{Characteristics of the PISN models in the present study. Explosion
  energy is in foe, where 1~foe\,=\,$10^{\,51}$~erg. All masses and mass
  coordinates are in solar masses. For some of the models featuring hydrogen
  down-mixing and up-mixing of $^{56}$Ni, the constant mass fractions
  of these elements are provided in parenthesis.} 
\label{table:model}
\begin{tabular}{|l|c|c|c|l|c|l|c|l}
\hline
model & Radius &M$_\mathrm{tot}$&M$_\mathrm{H}$&H-mix down                 &M$_\mathrm{Ni}$&Ni-mix up                  & E$_\mathrm{expl}$ & Notes\\
      & [\Rsun{}]& [\Msun{}]    &[\Msun{}]     &to M$_\mathrm{r}$ [\Msun{}]& [Msun{}]      &to M$_\mathrm{r}$ [\Msun{}]&[foe]              &   \\
\hline
150M      & 3450 & 93    & 4.9& --   & 0.03& 6 &  5 & original (Kozyreva et al. 2014)\\
150M-3000R& 3000 & 90    & 3.9& 30   & 1.18 & 6 & 9 & \\
150M-14foe& 2000 & 86    & 9.1& 10   & 1.31 & 5 & 14 & \\
150M-He   & 2000 & 86    & 1.6& --   & 1.27 & 5 & 14 & H$\rightarrow$He down to 30~\Msun{} M(He)=29~\Msun{}\\
150M-bf   & 2000 & 86    & 9.1& 10   & 1.31  & 5 & 9 & best-fitting\\
\hline
P150      & 1267 & 91    & 3.8& --   & 0.003& 4 & 6 & original (Gilmer et al. 2017)\\
P150ni10  & 1267 & 91    & 9.3& 10   & 1.37  & 10 (X$_\mathrm{Ni}$=0.14)& 9 & \\
P150ni30  & 1267 & 91    & 9.3& 10   & 1.37  & 30 (X$_\mathrm{Ni}$=0.045)& 9 & \\
P150ni50  & 1267 & 91    & 9.3& 10   & 1.4  & 50 (X$_\mathrm{Ni}$=0.015)& 9 & \\
P150-bf   & 1267 & 91    & 9.3& 10   & 1.37 & 1.6 & 9 & best-fitting\\
\hline
He90      & 18   & 90    & --  &    & 1.3 & 25 & 29 & original (Heger \& Woosley 2002)\\
He90-H001 & 18   & 90    & 1.6 & 10 (X$_\mathrm{H}$=0.01)& 1.3 & 25 & 29 & \\
He90-H01  & 18   & 90    & 7.3 & 10 (X$_\mathrm{H}$=0.1) & 1.3 & 25 & 29 & \\
He90-H02ni& 18   & 90    & 9.7 & 10--60 (X$_\mathrm{H}$=0.2)& 1.4 & 2 & 29 & centered $^{56}$Ni\\
          &      &       &     & $>$60 (X$_\mathrm{H}$=0.05) &  &  &  & \\
\hline
\end{tabular}
\end{table*}

Our models are based on two very massive non-rotating star models with M$_{\rm
ZAMS}$ = 150~\Msun{} at a metallicity \textit{Z}=0.001, evolved from the zero age main sequence (ZAMS) until the
start of the pair-instability phase with stellar evolution codes. The two
models, 150M and P150, were evolved with \textsc{bec}
\citep[][]{2007A&A...475L..19L,2014A&A...566A.146K} and \textsc{genec}
\citep[][]{2012A&A...537A.146E,2013MNRAS.433.1114Y},
respectively. Abundances in the original \textsc{GENEC} stellar
  evolution model P150 are scaled from solar metallicity ($Z=0.014$) down to
  $Z=0.001$ and alpha-enhanced in most cases, following
  observational constraints \citep{2003ApJ...591.1220L}.
Note that although both models have
the same total metallicity, it is distributed differently onto the various
elements. In particular, the calcium abundance is zero in 150M but $1.5 \times
10^{-6}$ in P150. 

Pair-instability explosions of the original models 150M and P150 were carried
out with \textsc{bec} and \textsc{flash}
\citep{2000ApJS..131..273F,2009arXiv0903.4875D,2013ApJ...776..129C,2015ApJ...799...18C},
respectively.
The models 150M and P150 have helium core masses of 73~\Msun{} and 83~\Msun{},
respectively. Both models build up equally massive CO-cores of 64~\Msun{}. In
the explosion, only small amounts of $^{56}$Ni are produced, namely
0.04\,\Msun{} in 150M and 0.003\,\Msun{} in P150.  Consequently, the resulting
light curves are dominated by cooling processes and do not show a peak powered
by $^{56}$Ni \citep{2014A&A...565A..70K,2017ApJ...846..100G}. Thus,
these models are not compatible with OGLE14-073 since its peak luminosity seems
to require about 1.3~\Msun{} of \,$^{56}$Ni.  

However, rotation-induced mixing would
effectively increases the CO-core mass.  Therefore, rotating siblings
of our 150~\Msun{} models would end up with higher mass CO-cores and,
in turn, yield more $^{56}$Ni in a PISN explosion.  For example, a
$Z=0.001$, 150~\Msun{} rotating model computed with \textsc{genec}
develops an 85~\Msun{} helium core and an 80~\Msun{} CO-core
(R.\,Hirschi, private communication,
\citealt{2013MNRAS.433.1114Y}).  Besides that, more massive stars at
slightly lower metallicity ($Z<0.001$) retain the hydrogen-helium
envelope similar to 150M or P150 (R.\,Hirschi, N.\,Yusof, private
communication).  Therefore, we decided that using 150M and P150 as
reference models is a reasonable starting point.  

Ideally, a large
grid of models would be computed, with initial masses between 150 and
200 \Msun ($^{56}$Ni yield from 0.003~\Msun{} up to 13~\Msun{}), with
varying initial metallicities and different degrees of rotation to
find the best match to OGLE14-073. However, computing the evolution of
very massive star models is very challenging due to the loose
gravitational binding of the hydrogen-rich envelope and the proximity
to the Eddington limit. Furthermore, mass loss is uncertain in these
stars and they may experience large, $\eta$-Carinae-like, mass-loss
episodes \citep{2015ASSL..412...77V}. Given all these complications,
we adopt a reverse-engineering approach in this study to find
out which modifications of the original models would enable a fit to
the observed properties of OGLE14-073. The modifications made to the
original models are given in Table\,\ref{table:model} and we add
suffixes to the original model names to identify them. We explain
below in more detail the various modifications made and the
underlying motivations.

In the original observational paper by \citet{2017NatAs...1..713T}, the authors
find a number of similarities between OGLE14-073 and SN~1987A. In particular,
H$\alpha$ is observed during the whole observational period. Hence, the
chemical structure of the OGLE14-073 ejecta might resemble that of SN~1987A.
Studies of SN~1987A conclude that hydrogen was mixed deep into the ejecta and
that $^{56}$Ni was mixed into ejecta regions far away from the centre
\citep{1993A&A...270..249U}. Such a mixing pattern has been confirmed in
multi-dimensional simulations of core collapse SNe
\citep[e.g.][]{2015A&A...577A..48W}. Similar mixing processes might be active
in PISN. Indeed, 2.5-D simulations of rotating PISNe by
\citet{2013ApJ...776..129C} show the formation of Rayleigh--Taylor
instabilities and asymmetries that lead to mixing in the ejecta.  Based on this and the
similarities between OGLE14-073 and SN\,1987A, we explored whether the presence
of hydrogen in OGLE14-073 might be explained by artificially mixing hydrogen
deep into the ejecta in both explosion models. We found that modified versions
of models 150M and P150 with a hydrogen mass fraction of $0.1$ down to a mass
coordinate M$_r=10$\,\Msun{} produce bolometric light curves in good agreement
with OGLE14-073.

\begin{figure*}
\centering
\includegraphics[width=\columnwidth]{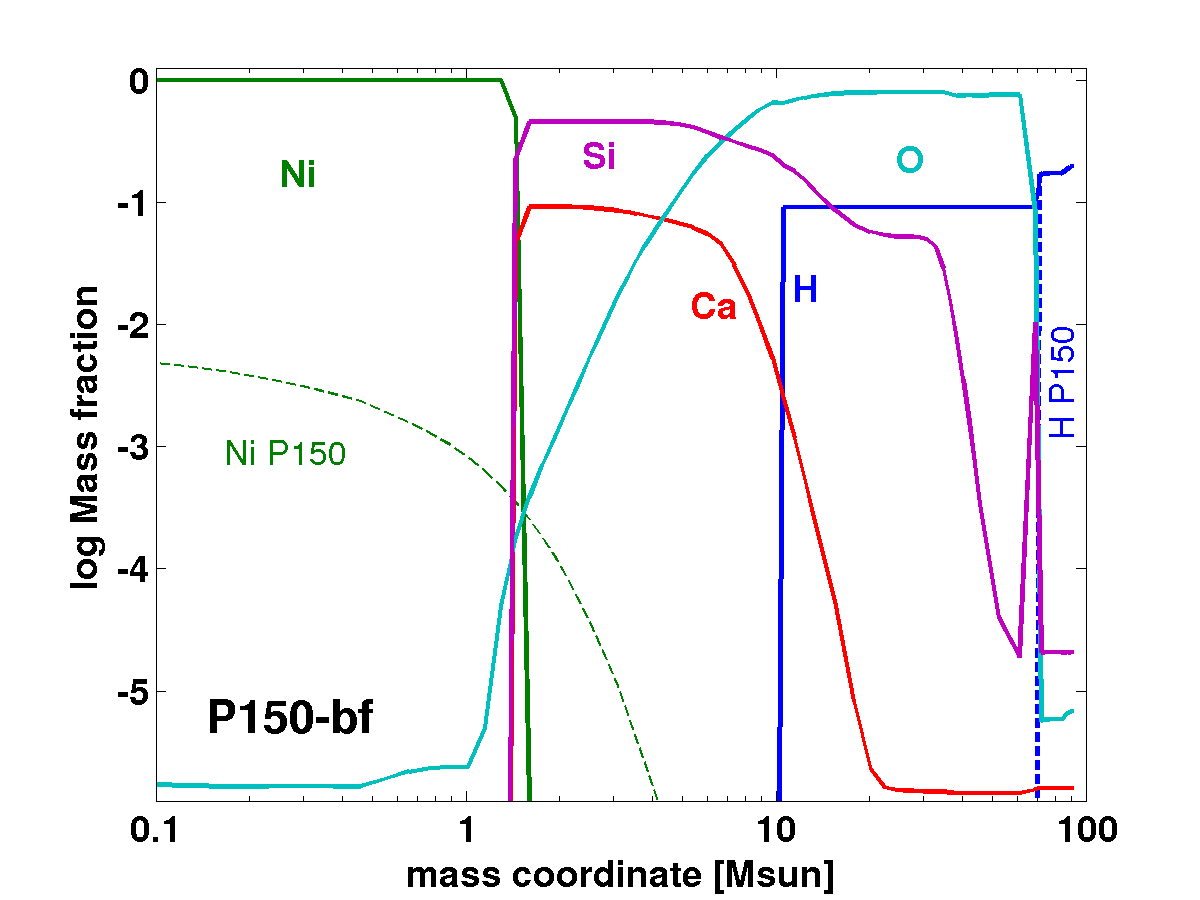}~
\includegraphics[width=\columnwidth]{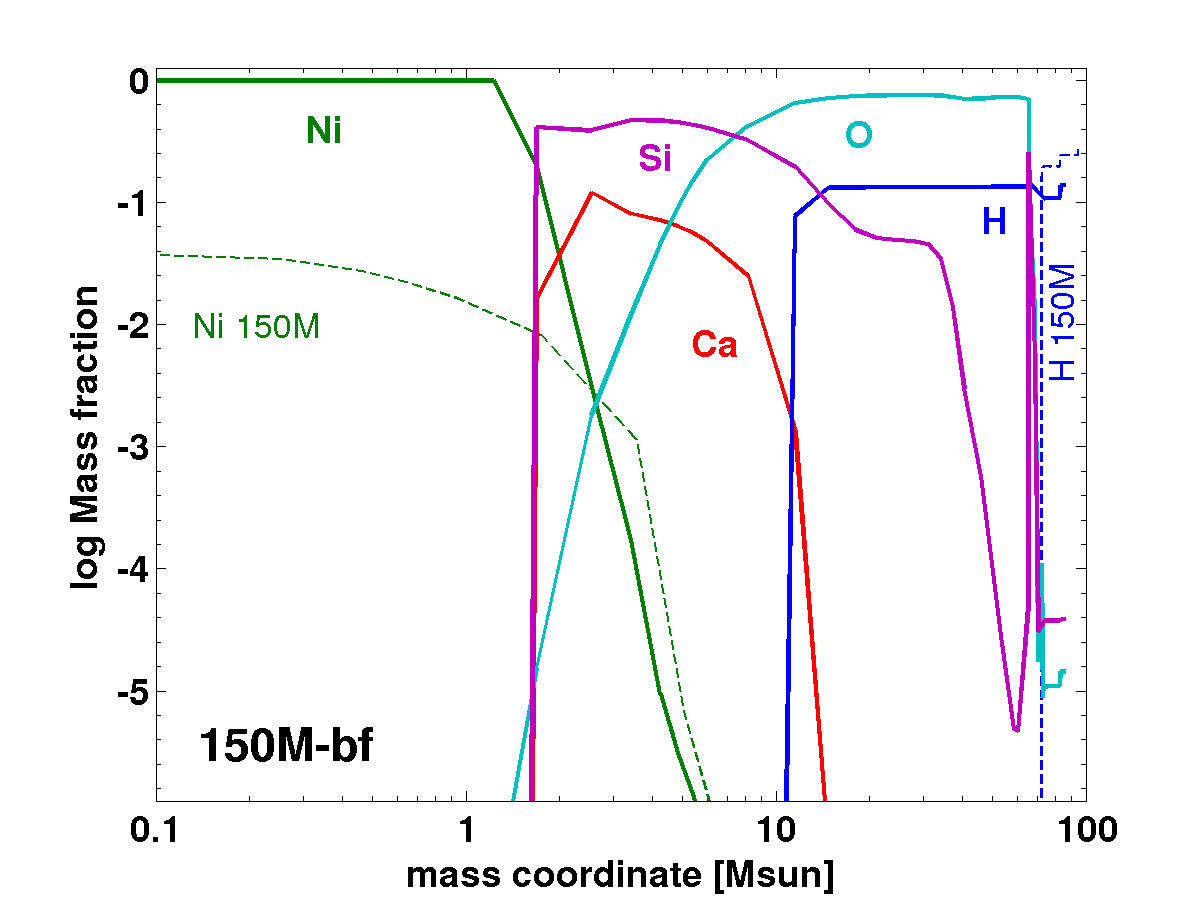}
\caption[Distribution of selected species (hydrogen, oxygen, silicon, calcium, and $^{56}$Ni in the modified models P150-bf (left) and 150M-bf (right)]{
Distribution of selected species (hydrogen, oxygen, silicon, calcium, and $^{56}$Ni in the modified models P150-bf (left) and 150M-bf (right). Dashed lines represent the distribution of hydrogen and $^{56}$Ni in the original models   P150 and 150M. }
\label{figure:chemie}
\end{figure*}

The best-fitting models of SN~1987A have $^{56}$Ni mixed throughout
the entire ejecta \citep{1988PASAu...7..490N}. Thus, we also explored
the effect of mixing $^{56}$Ni outward in the model P150. We
constructed models, in which $^{56}$Ni was smeared out
uniformly inside the inner 10, 30 and 50 solar masses, but found that
this mixing does not improve the match between the resulting synthetic
light curves and OGLE14-073 (see discussion in
Section~\ref{sect:results}).

Finally, we lowered the hydrogen-to-helium mass fraction ratio in the
outer shells of the model 150M (H:He$\simeq$1:4 in the original model,
while we set H:He$\simeq$1:8), to reduce the high cooling emission
after the shock breakout.  For the same reason, i.e.\ to reduce the
strong cooling signature, model 150M was truncated at a radius,
$R=2000$\,\Msun{} (the original radius of 150M is 3475\,\Msun{}). In
Figure~\ref{figure:chemie}, we show the radial distribution of 
hydrogen, $^{56}$Ni and a few other selected chemical species for our 
best-fitting models 150M-bf and P150-bf, and compare them to the distribution 
of hydrogen and $^{56}$Ni in the original models 150M and P150.

\subsection[Light curve modelling]{Light curve modelling}
\label{subsect:method2}

We follow the supernova explosion and model the light curve
with the one--dimensional multigroup radiation hydrodynamics code
\textsc{stella}
\citep[][]{2006A&A...453..229B,2014A&A...565A..70K}. We mapped the
models into \textsc{stella} when the shock is at the bottom (models
150M) or in the middle (models P150) of the H-He envelope. The energy
of the pair-instability explosion (in the modified models) was
increased by multiplying the velocity of the basic model (150M and
P150) by a factor 1.3. In the present simulations, we use 100 frequency
bins.  The opacity includes photoionization, free-free absorption,
electron scattering processes assuming local thermodynamical equilibrium in the
plasma, and line interactions. The line opacity is calculated using a
database of about 160,000 spectral lines from \citet{Kurucz1995a} and
\citet{1996ADNDT..64....1V}. Energy deposition from $^{56}$Ni and $^{56}$Co 
radioactive decay is treated in a one-group diffusion approximation
according to \citet{1995ApJ...446..766S}.

When calculating opacities, \textsc{STELLA} treats only a limited set
of species, which generally play an important role in SNe. These are H, He, C,
N, O, Ne, Na, Mg, Al, Si, S, Ar, Ca, stable Fe, which represents heavy
iron-group elements, radioactive $^{56}$Co and stable Ni and radioactive
$^{56}$Ni. We note that ejecta models extracted from \textsc{STELLA}, and used
for the spectral synthesis post-processing described in the next Section, are
limited to the same set of species.

\subsection[Spectral synthesis]{Spectral synthesis}\label{subsect:spectra}

To obtain spectral time series for the models P150-bf and 150M-bf, we
use the Monte Carlo radiative transfer code \textsc{artis}
\citep{2009MNRAS.398.1809K}.  Using a detailed wavelength-dependent opacity
treatment, \textsc{artis} solves the time-dependent radiative transfer problem
in homologously expanding supernova ejecta. This includes a detailed simulation
of the propagation of $\gamma$-ray photons and ultraviolet-optical-infrared
photons. For $\gamma$-ray photons \textsc{artis} accounts for interactions with
matter by Compton scattering, photo-electric absorption and pair production
\citep{sim2008a}. For ultraviolet-optical-infrared photons \textsc{artis}
accounts for electron scattering, free-free, bound-free and line
interactions. The latter are treated in the Sobolev approximation
\citep{sobolev1957a} using a generalised scheme that enforces statistical
equilibrium and allows for a parameter-free treatment of line fluorescence
\citep{lucy2002a,lucy2003a}. The ionisation and thermal balance equations are
solved self-consistently with the radiative transfer problem. Excitation is
treated approximately by assuming local thermodynamic equilibrium.

As input model for our \textsc{artis} simulations we use the density and
abundance structure from \textsc{stella} at 10\,d after explosion, when the
ejecta approach homologous expansion. In \textsc{artis} this ejecta structure
is mapped to a $100^{\,3}$ uniform Cartesian grid and follows homologous
expansion. On this grid, we then simulate the evolution of the radiation field
by propagating $4\times10^{\,7}$\, Monte Carlo quanta for 150
logarithmically-spaced time steps from 50 to 450\,d past explosion.  Once the
Monte Carlo quanta escape from the simulation domain they are binned in time
and on a logarithmic wavelength grid spanning 1,000 bins from 375 to
30,000\,\AA\ to obtain the spectral time series.
\begin{figure}
\centering
\includegraphics[width=\columnwidth]{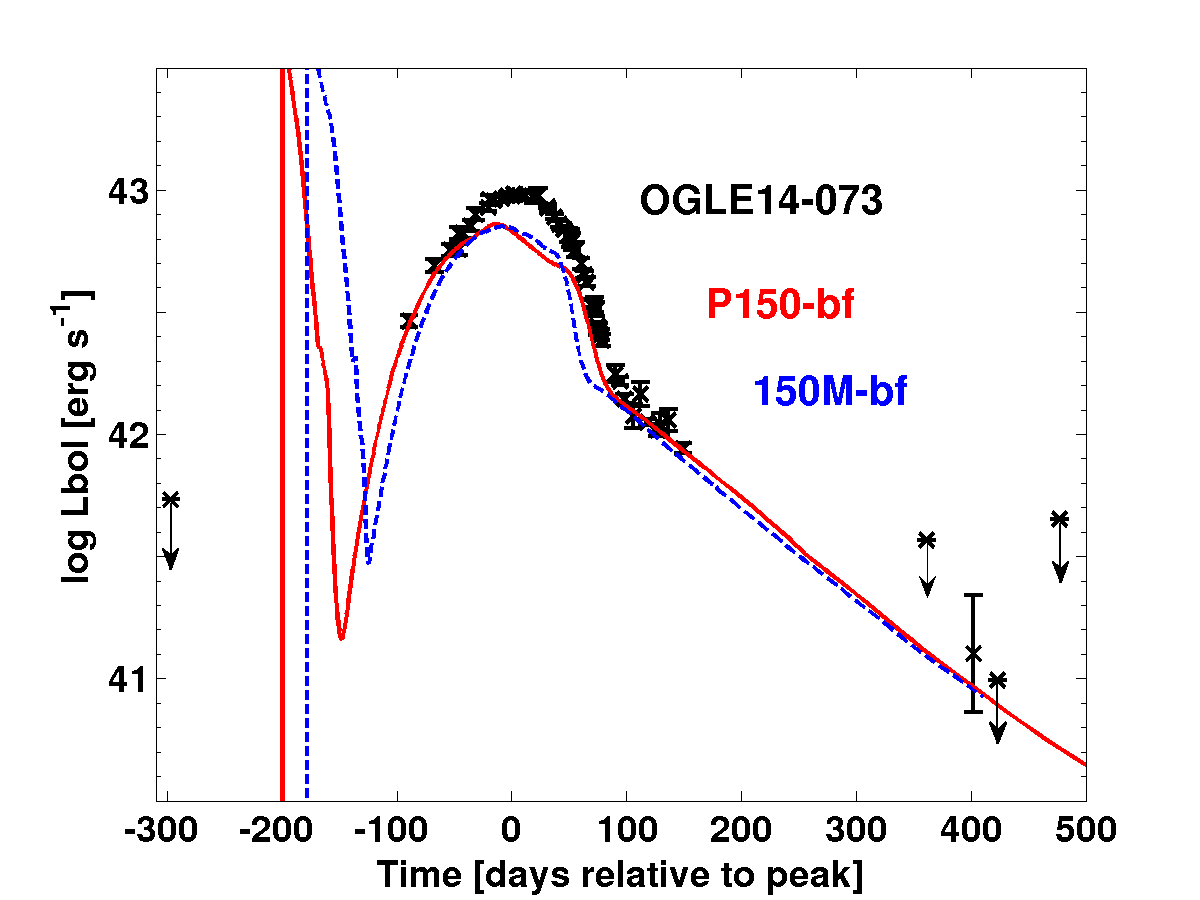}
\caption[Bolometric light curves]{Bolometric light curve of OGLE14-073 (black
  crosses), compared to the synthetic \textsc{stella} light curves for models
  150M-bf (blue dashed) and P150-bf (red solid).}
\label{figure:Lbol}
\end{figure}

\begin{figure*}
\centering
\includegraphics[width=\columnwidth]{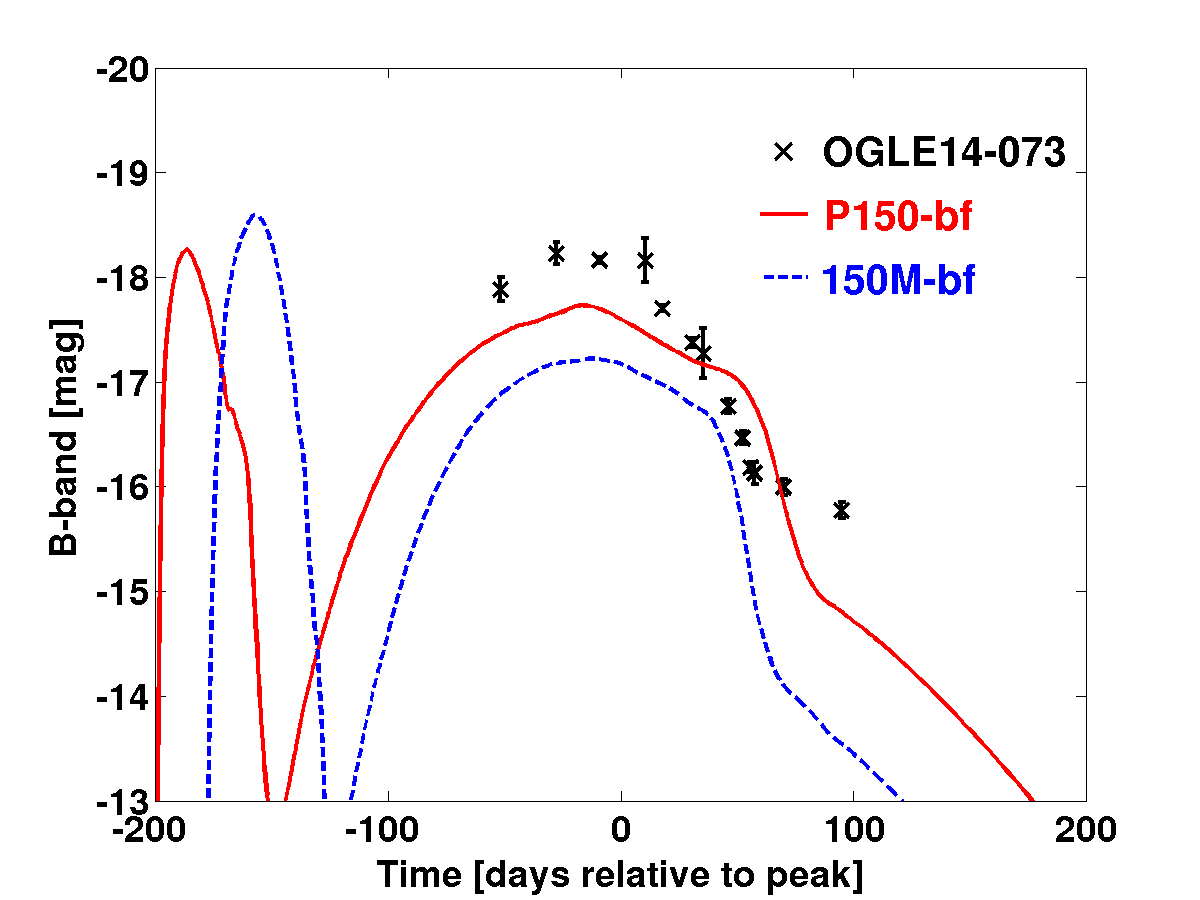}~
\includegraphics[width=\columnwidth]{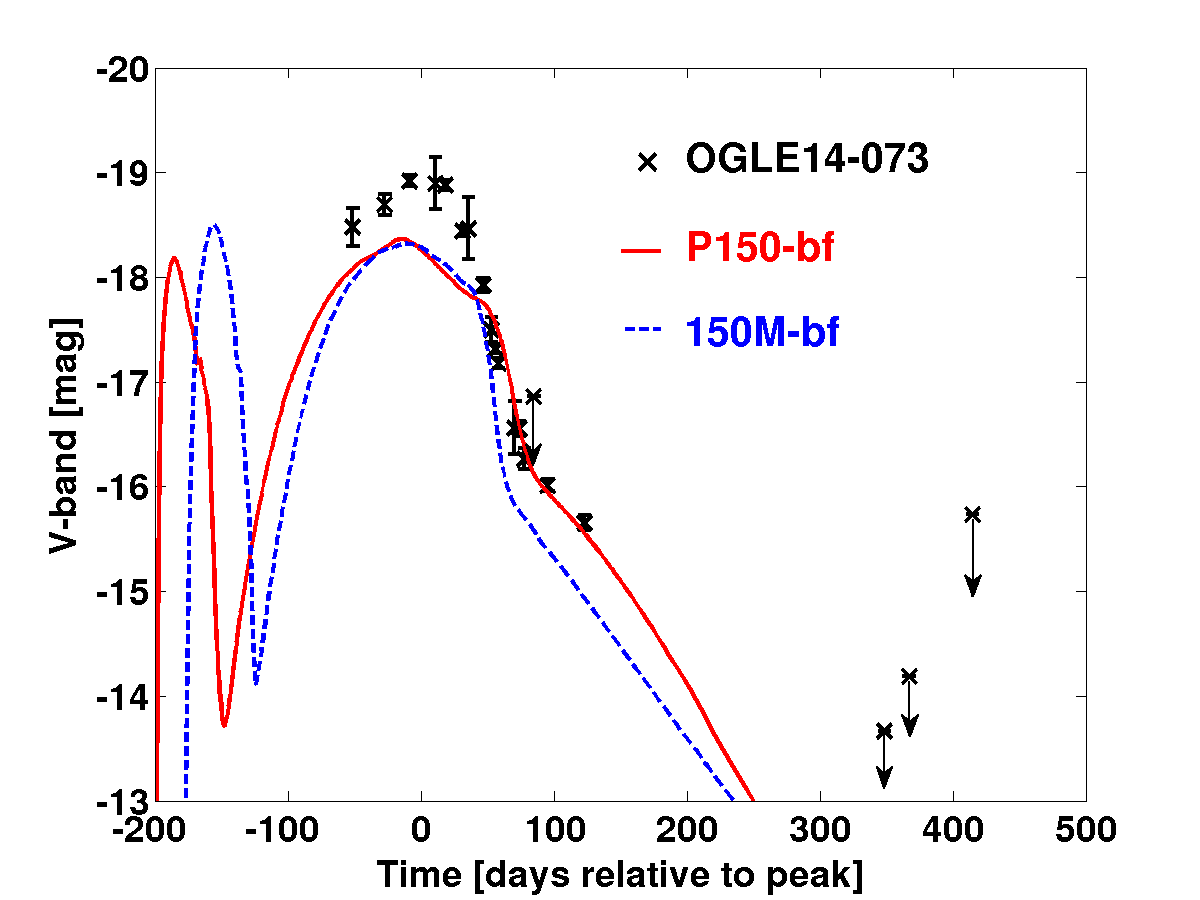}\\
\includegraphics[width=\columnwidth]{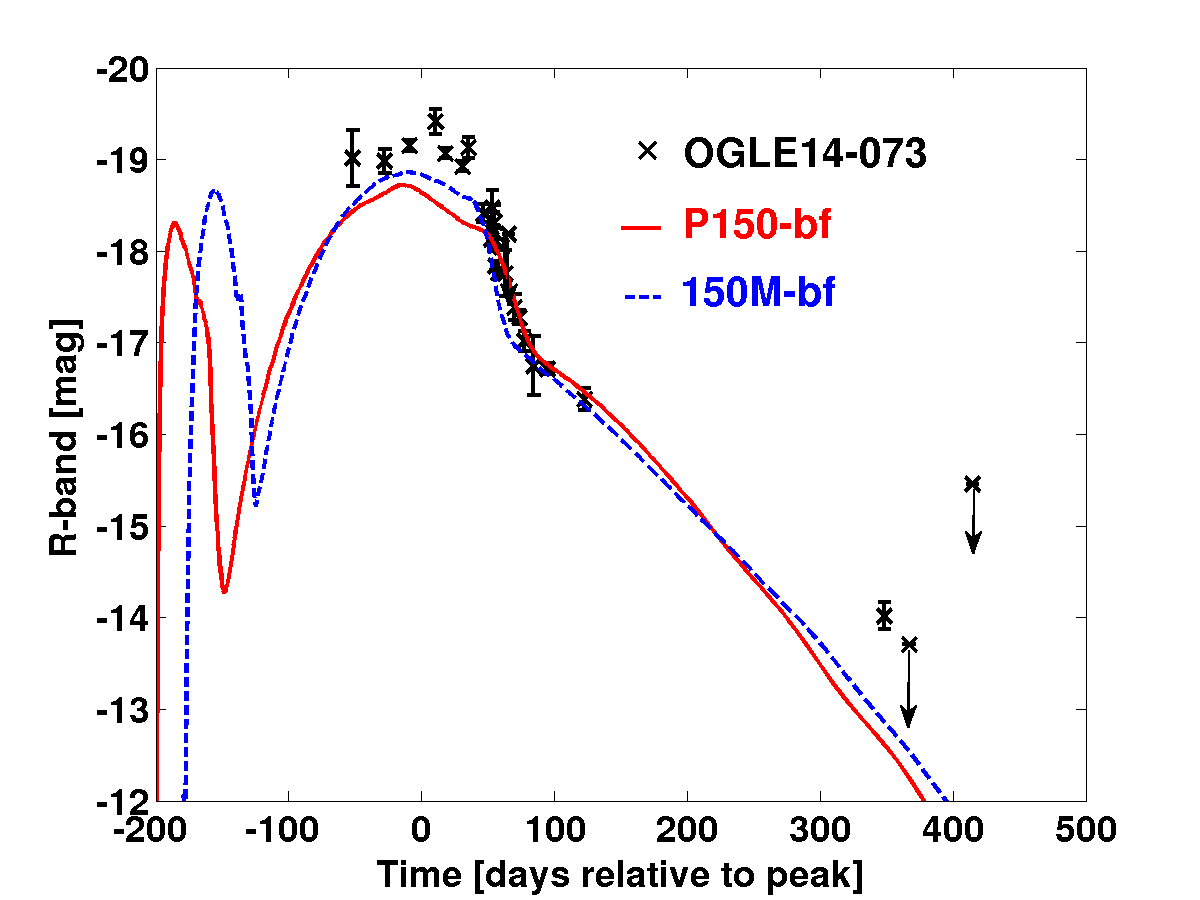}~
\includegraphics[width=\columnwidth]{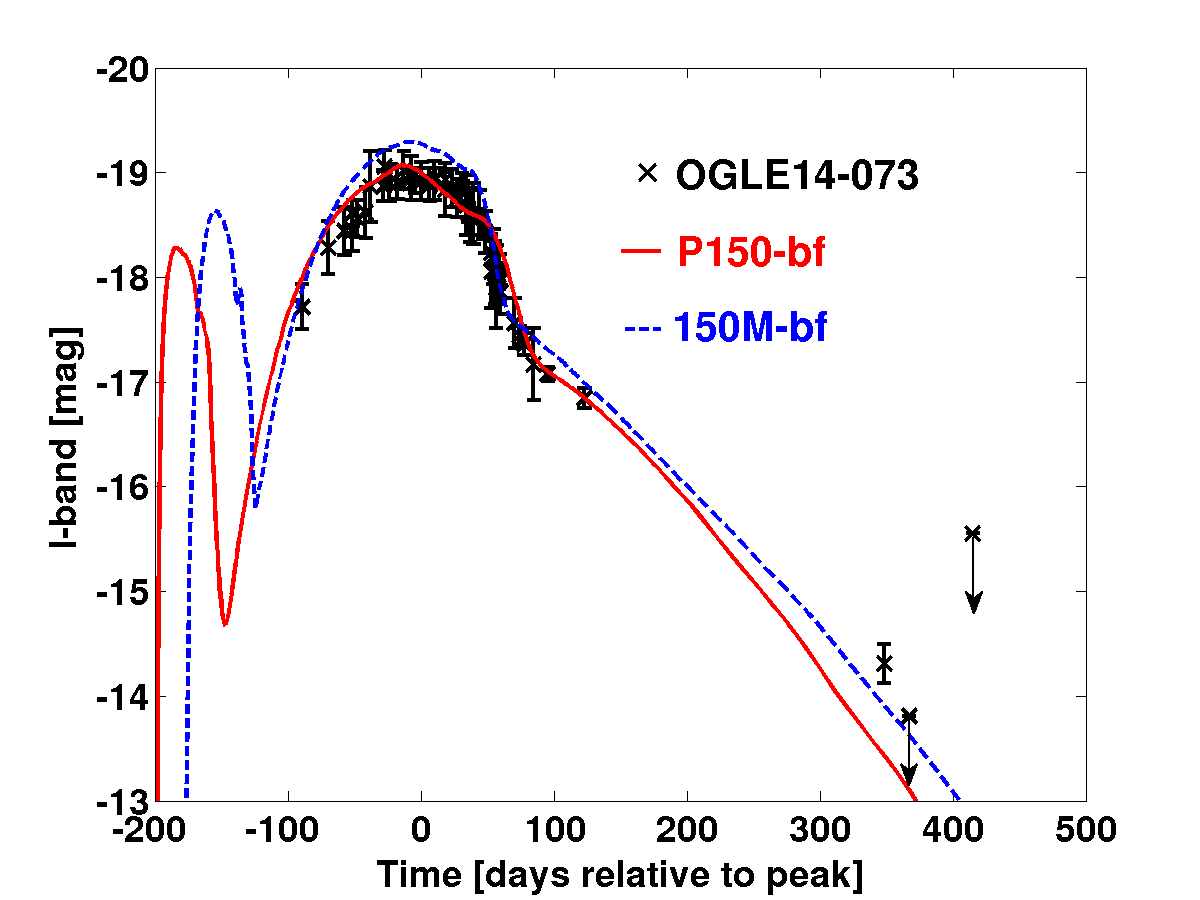}
\caption[Light curves in broad bands]{Light curves of OGLE14-073 (black
  crosses), compared to synthetic \textsc{stella} light curves for models 150M-bf (blue dashed) and P150-bf (red solid) in the $B$
  (upper left), $V$ (upper right), $R$ (lower left) and $I$ (lower right)
  filter bands.}
\label{figure:bands}
\end{figure*}

For our simulation we use an atomic data set that is based on the cd23\_gf-5
model of \citet{2009MNRAS.398.1809K} but extended \textbf{by} H
lines which were not part of the original compilation. The cd23\_gf-5
data set is based on the atomic line list by \citet{Kurucz1995a}. A
grey approximation is used in optically thick cells for $t < 105$\,d
to speed up the calculations \citep[cf.][]{2009MNRAS.398.1809K}. Local
thermodynamic equilibrium is assumed for $t < 58$\,d. At later epochs,
we use the detailed ionisation treatment of \textsc{artis} to solve
the ionisation balance as described by
\citet{2009MNRAS.398.1809K}. For the excitation balance we follow a
different approach and adopt a nebular approximation
\begin{equation}
  \frac{n_{i,j,k}}{n_{0,j,k}}=W \frac{g_{i,j,k}}{g_{0,j,k}}\, \exp\left( -\frac{\epsilon_{i,j,k}-\epsilon_{0,j,k}}{k_{\mathrm{B}}T_{\mathrm{R}}}\right) 
  \label{eq:nebapprox}
\end{equation}
with symbols as defined in \citet{2009MNRAS.398.1809K}. For metastable levels
(i.e.\ levels without permitted electric dipole transitions to the
ground state) the dilution factor $W$ is set to $1$.

\section[Results and Discussion]{Results and Discussion}\label{sect:results}

\subsection[Bolometric properties and broad band magnitudes]{Bolometric properties and broad band magnitudes}
\label{subsect:broad}

In Figures~\ref{figure:Lbol} and \ref{figure:bands}, we show our best-fitting
synthetic light curves from the \textsc{stella} calculations for our modified
models 150M-bf and P150-bf. Bolometric light curves and the evolution in the
\textit{B}, \textit{V}, \textit{R}, and \textit{I} bands are shown,
respectively. The models qualitatively follow the observed light-curve
evolution of OGLE14-073. In particular, the slope during the rise to the peak,
the smooth dome-like shape of the peak phase, the drop after the peak and
transition to the tail are well-reproduced. The bolometric luminosity at peak
is underestimated by 0.15~dex compared to OGLE14-073. The light curves are
redder by 0.9~mags and 0.4~mags in \textit{B} for 150M-bf and P150-bf
respectively. In \textit{V} and \textit{R}, both models are too red by 0.3~mags
while a perfect match between models and observations is achieved in the
\textit{I} band. This colour effect could be related to the LTE assumption in
\textsc{stella} or its simplified treatment of line opacities. For other PISN
models it was suggested that the red colours result from line blanketing due to
the high metal content of the models, particularly of \,$^{56}$Ni
\citep{2013MNRAS.428.3227D}. As $^{56}$Ni is confined to the central ejecta
regions in our model, we expect the effect of line blanketing to be much
weaker. In fact, our synthetic spectra presented in
Section~\ref{subsect:spectrares} are in good agreement with the broad-band SED
observed in OGLE14-073.

\begin{figure}
\centering
\includegraphics[width=\columnwidth]{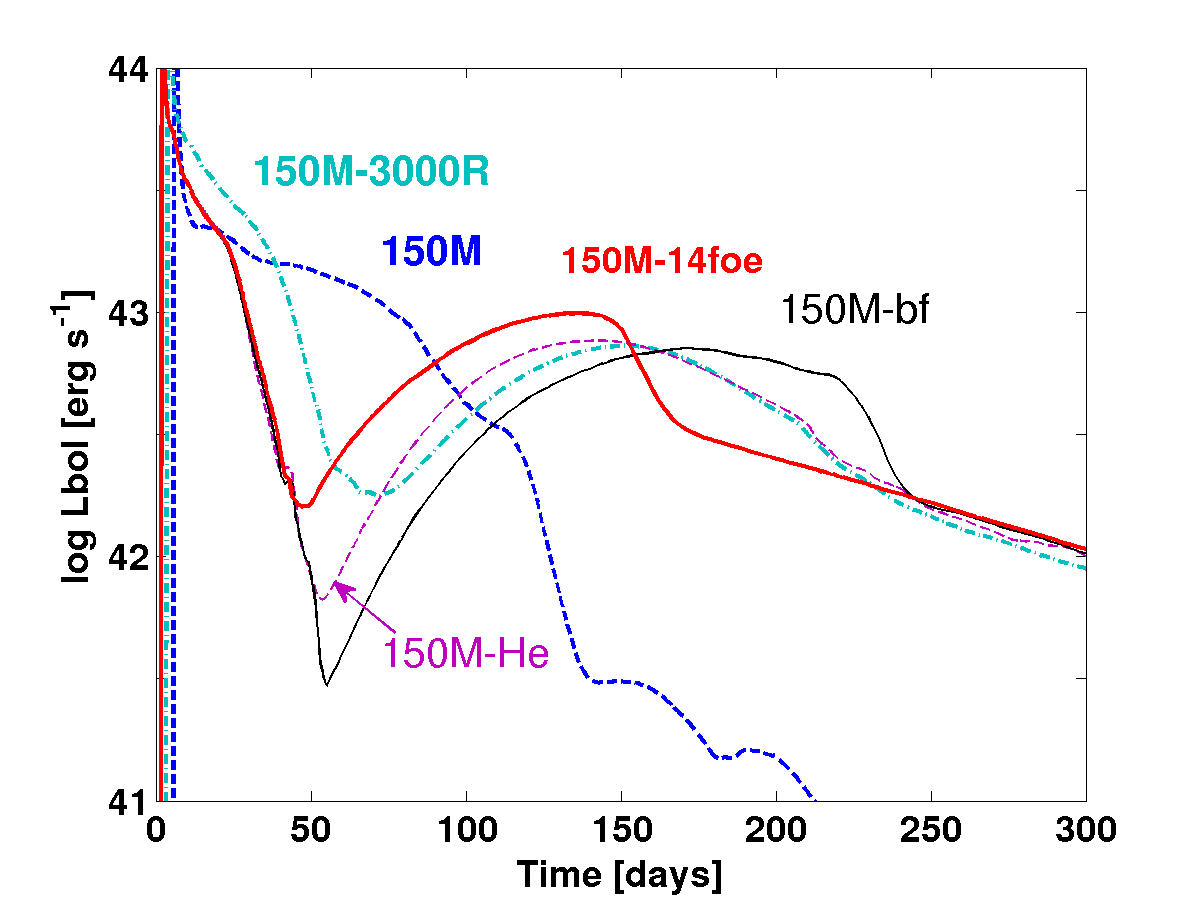}
\caption[Bolometric light curves for the set of 150M models with
different explosion energy, radius, hydrogen fraction]{Bolometric
  light curves for the set of 150M models with different explosion
  energies, radii and hydrogen fraction.}
\label{figure:150Mvary}
\end{figure}

\begin{figure}
\centering
\includegraphics[width=\columnwidth]{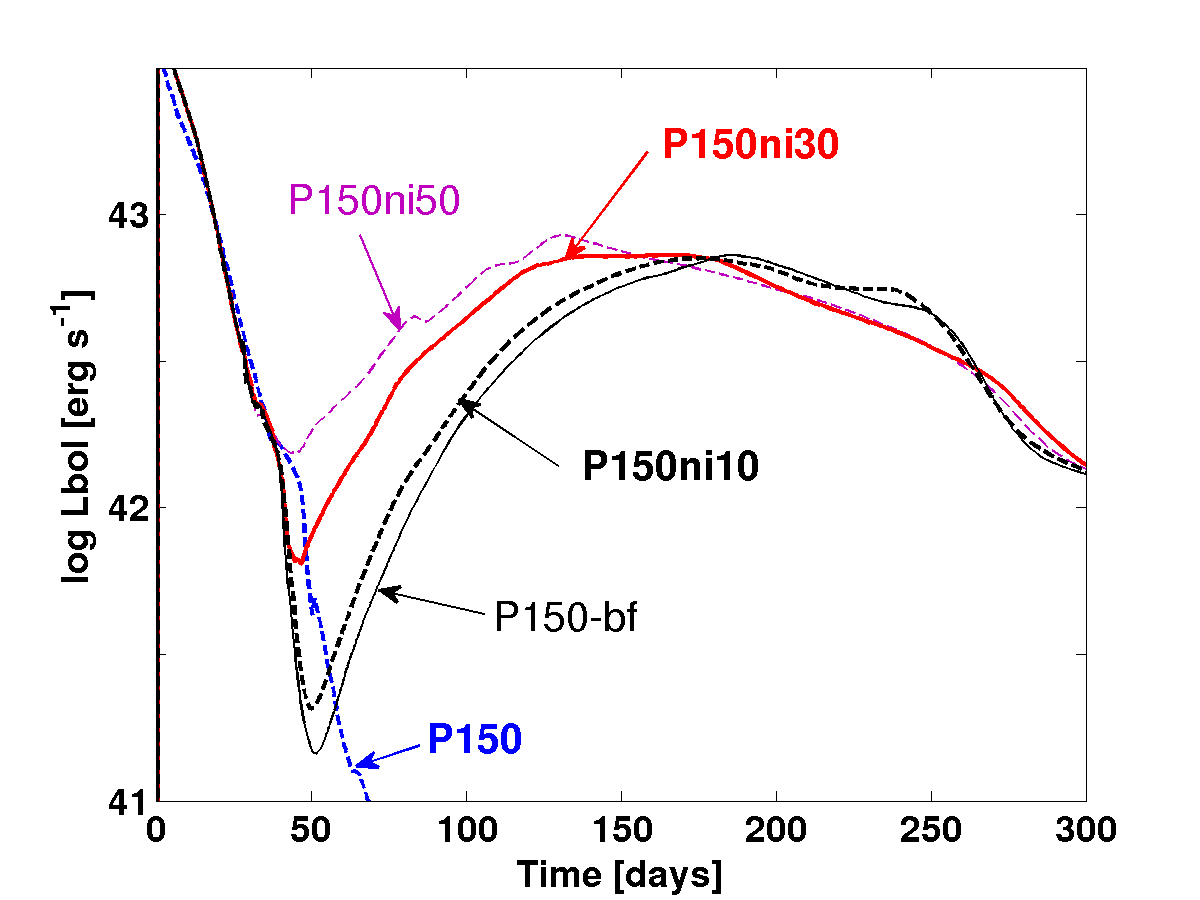}
\caption[Bolometric light curves for the set of P150 models showing
the influence of an extended $^{56}$Ni distribution]{Bolometric light curves
  for the set of P150 models showing the influence of an extended $^{56}$Ni
  distribution: the original model P150 (dashed), the best-fitting
  modified model P150-bf (thin), and the models P150ni10 (thick
  dashed), P150ni30 (thick), and P150ni50 (thin dashed) with $^{56}$Ni
  distributed up to $M_\mathrm{r}=10, 30$, and 50~\Msun{},
  correspondingly.}
\label{figure:P150vary}
\end{figure}

\begin{figure}
\centering
\includegraphics[width=\columnwidth]{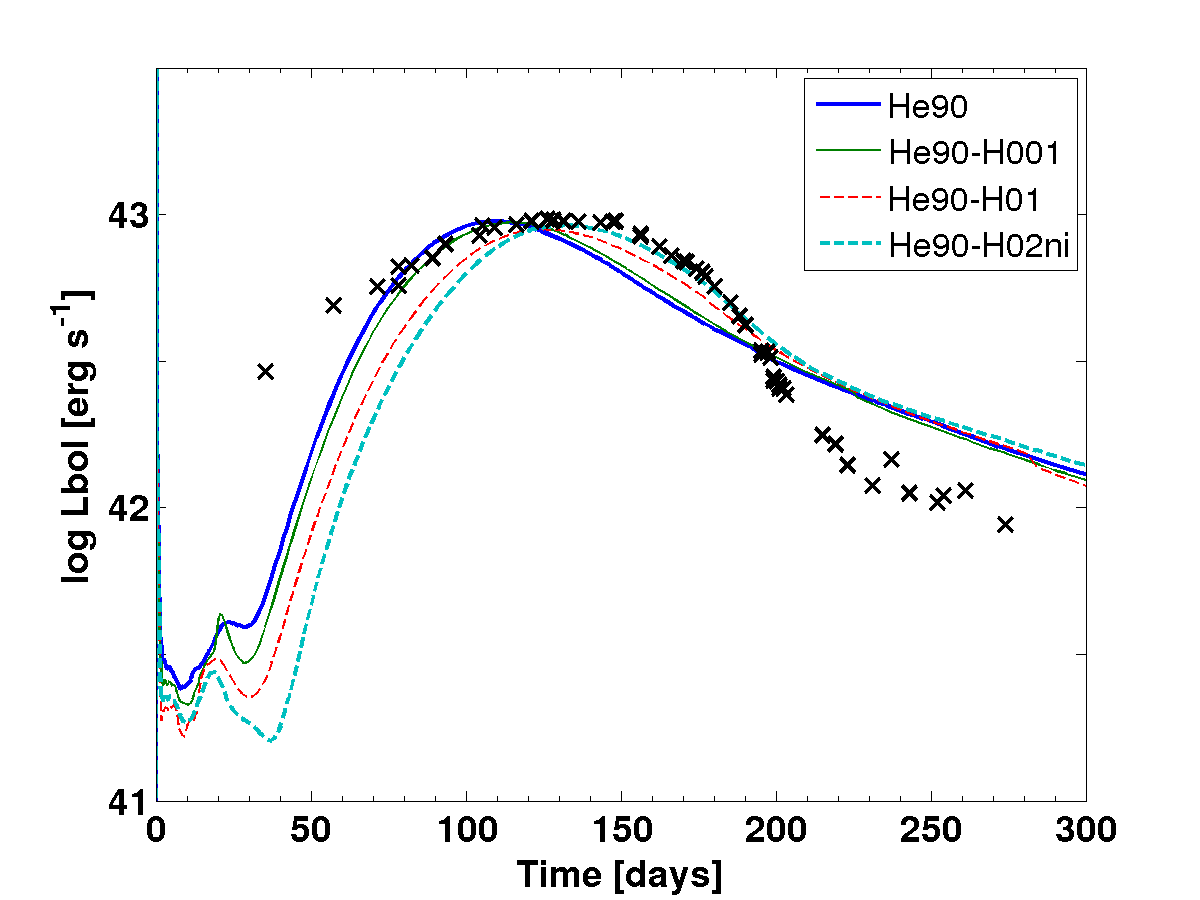}
\caption[Bolometric light curves for the set of P150 models showing the
influence of the extended $^{56}$Ni distribution]{Bolometric light curves for the set of He90 models: He90 (solid thick, original), He90-H001 (solid thin,
  X(H)=0.01), He90-H01 (X(H)=0.1), and He90-H02ni (X(H)=0.2, $^{56}$Ni is
  confined to inner 2~\Msun{}). The data points of OGLE14-073 (black crosses)
  are shifted by  35~days (relative to the first detection). See details in
  Table~\ref{table:model}.}
\label{figure:he90vary}
\end{figure}

In the following, we present our findings when exploring the various
modifications of the original PISN models, in particular with respect
to the bolometric light curve. Specifically, we investigate the
consequences of changes of the amount of $^{56}$Ni and hydrogen
mixing, varying the explosion energy and the radius of the
progenitor. The resulting trends are illustrated in
Figures~\ref{figure:150Mvary}, \ref{figure:P150vary}, and
\ref{figure:he90vary}:

\begin{itemize}
\item Larger radius (model 150M-3000R) and higher energy (150M-14foe) for the
  model 150M increase the luminosity during the cooling phase after shock
  breakout and extend its duration as shown in Figure~\ref{figure:150Mvary}.
  As there are no data between the non-detection limit (160~days before
  maximum) and the first discovery observation, the cooling phase cannot be
  well constrained.
\item A higher explosion energy reduces the diffusion time
  ($t_\mathrm{diff}\,\sim\,1/E$, \citealt{1982ApJ...253..785A}), hence,
  the light curve peaks earlier. We demonstrate this effect with the
  model 150M-14foe which is the same as 150M-bf but
  with the explosion energy increased by 5~foe (see
  Figure~\ref{figure:150Mvary}).  150M-14foe peaks 50~days earlier
  than our best fitting model 150M-bf.
\item The deeper hydrogen is mixed into the ejecta, the broader the
  dome-like maximum phase is. Hydrogen remains the main element
  providing electrons and contributing to the total opacity via
  electron scattering. Therefore, substituting hydrogen with helium
  leads to a sharper rise to peak. We demonstrate the He-effect
  with model 150M-He (\textbf{see} Figure~\ref{figure:150Mvary}).
\item Mixing of $^{56}$Ni shortens the effective diffusion time and
  reduces the time between explosion and maximum light. We run model P150
  with $^{56}$Ni distributed uniformly in the innermost 10~\Msun{}
  (model P150ni10), 30~\Msun{} (model P150ni30) and 50~\Msun{} (model
  P150ni50), and show the resulting light curves in
  Figure~\ref{figure:P150vary}. The resulting light curves have a shallower
  minimum after post shock breakout cooling and in turn start rising
  earlier than the model P150-bf with centrally concentrated
  nickel. On top of that, the dome-like light curve loses the
  ``OGLE14-073'' shape and tends to peak earlier, while the decline
  after the peak becomes flattened. We conclude that an extended
  distribution of $^{56}$Ni makes the model less applicable to
  OGLE14-073.
\end{itemize}

To explore whether more compact models may improve the overall match between the synthetic and observed bolometric light curve, 
we carry out additional simulations based on the PISN model He90 from \citet{2002ApJ...567..532H}. He90 is a compact
(18~\Rsun{}) helium star model which produces 1.3\,\Msun{} of $^{56}$Ni in the
pair-instability explosion. Peak luminosity is reached 60~days earlier (about
100 days after the explosion) than for the hydrogen-rich models 150M-bf and P150-bf,
as shown in Figure~\ref{figure:he90vary}. At the same time, the overall light
curve becomes narrower and does not reproduce the slopes of the OGLE14-073
light curve. We test the model He90 with a modified chemical composition, i.e.\
with hydrogen mixed down to mass coordinate M$_\mathrm{r}=10$\,\Msun{}. The
models He90-H001, He90-H01, and He90-H02ni have different hydrogen mass
fraction: 0.001, 0.1, and 0.2, correspondingly. We kept the original
distribution of $^{56}$Ni (it is naturally distributed in 20~\Msun{}) in the
models He90, He90-H001, and He90-H01. In the model He90-H02ni, $^{56}$Ni is
located in the central 2~\Msun{} to attempt to extend the rising phase of the
light curve. Nevertheless, these
modifications did not improve the match. We conclude that the compact helium
model He90 and its modifications do not result in observational properties
that match OGLE14-073.

Before shifting our focus towards the spectral properties of OGLE14-073, we
briefly summarize our findings based on modelling the
bolometric and broad-band light curves. By modifying two first-principle PISN models, we are able to produce light curves that
qualitatively reproduce the observed evolution of OGLE14-073. A crucial
ingredient is to invoke down-mixing of hydrogen into the inner ejecta regions,
similar to what has been observed for SN~1987A. As anticipated in the study of
\citet{2017NatAs...1..713T}, a fairly high $^{56}$Ni mass is needed to achieve a
peak brightness similar to OGLE14-073. Unlike for SN~1987A, we do not require
to mix radioactive $^{56}$Ni from the centre to the outer ejecta regions.
Instead, our best-fit models have centrally located $^{56}$Ni. Finally, the
explosion energy and progenitor radius are somewhat unconstrained since they
mainly affect the cooling phase, for which there is little observational
information.

\subsection[Spectral properties]{Spectral properties}\label{subsect:spectrares}

\begin{figure*}
\centering
\includegraphics[width=\columnwidth]{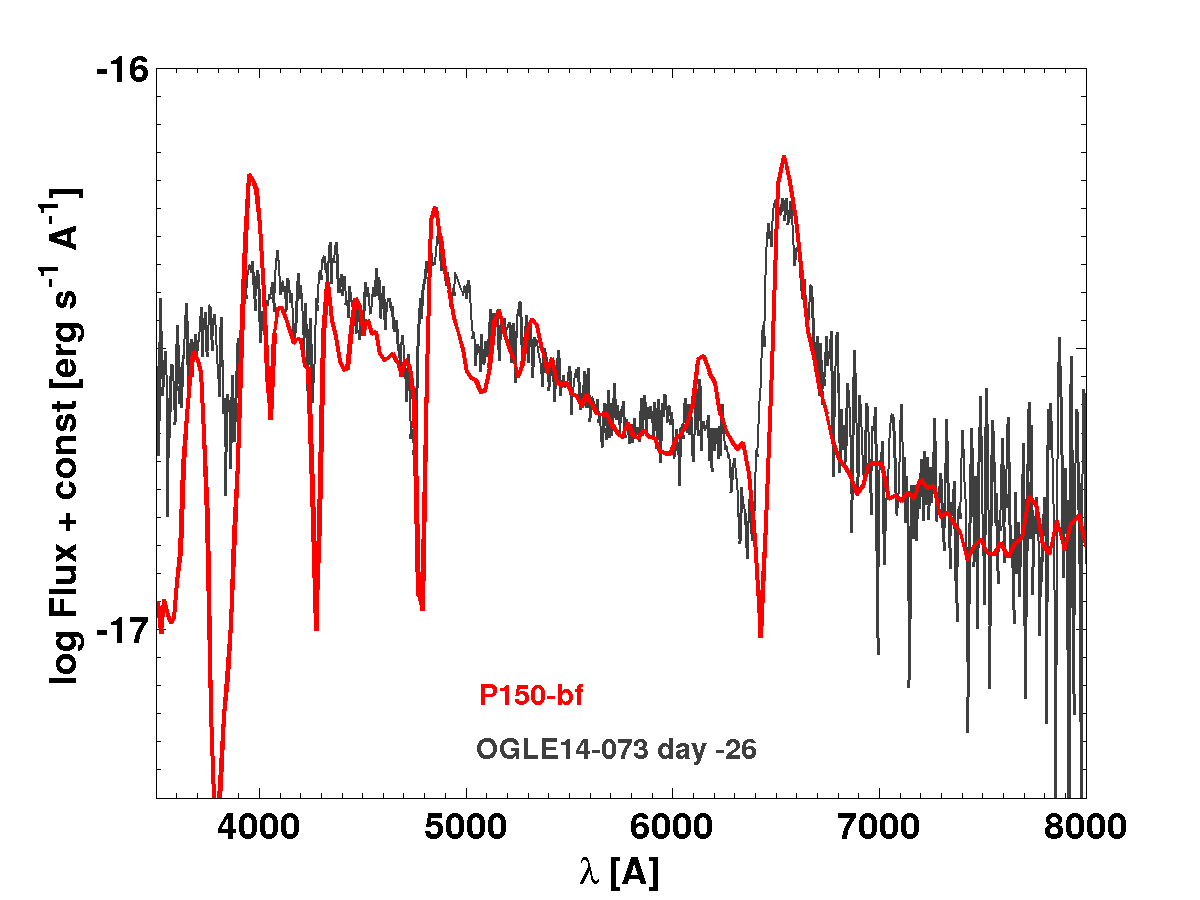}~
\includegraphics[width=\columnwidth]{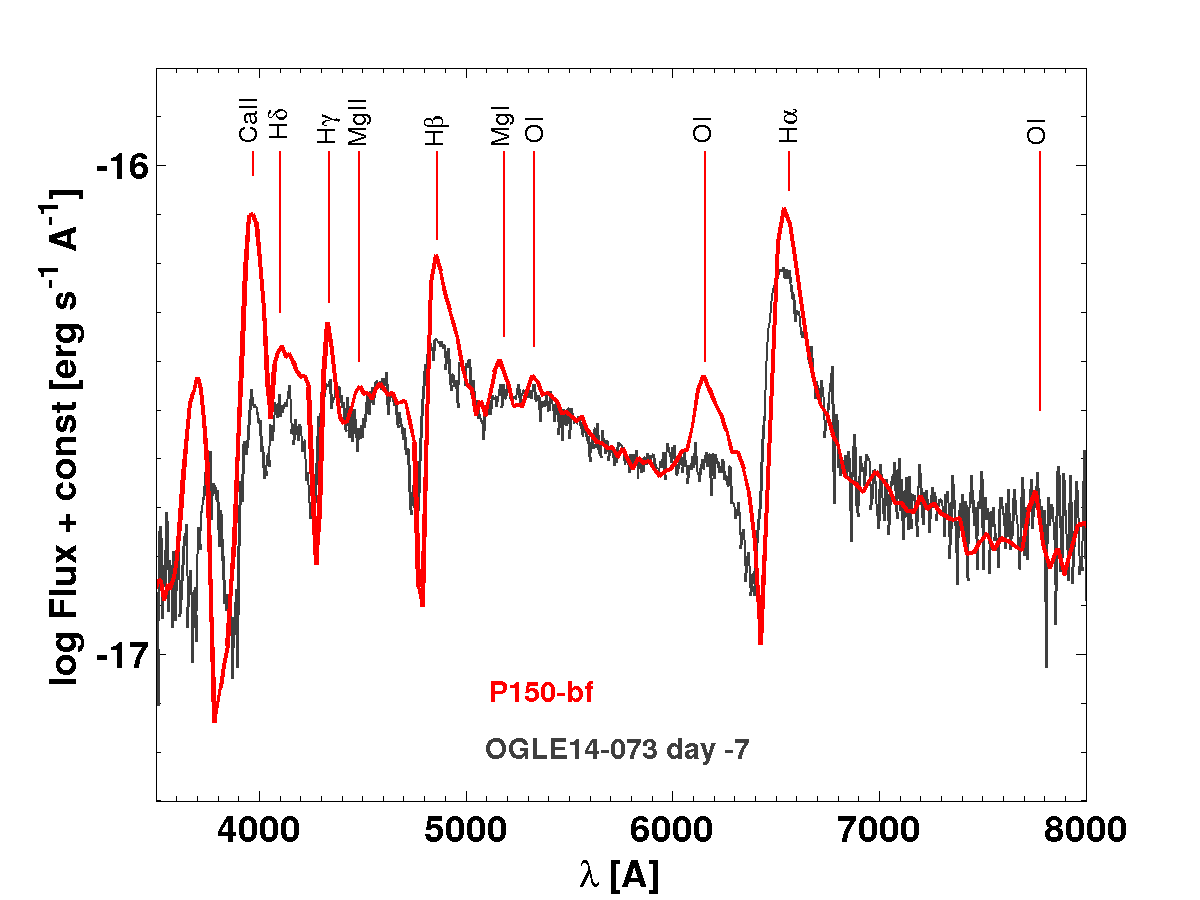}\\
\includegraphics[width=\columnwidth]{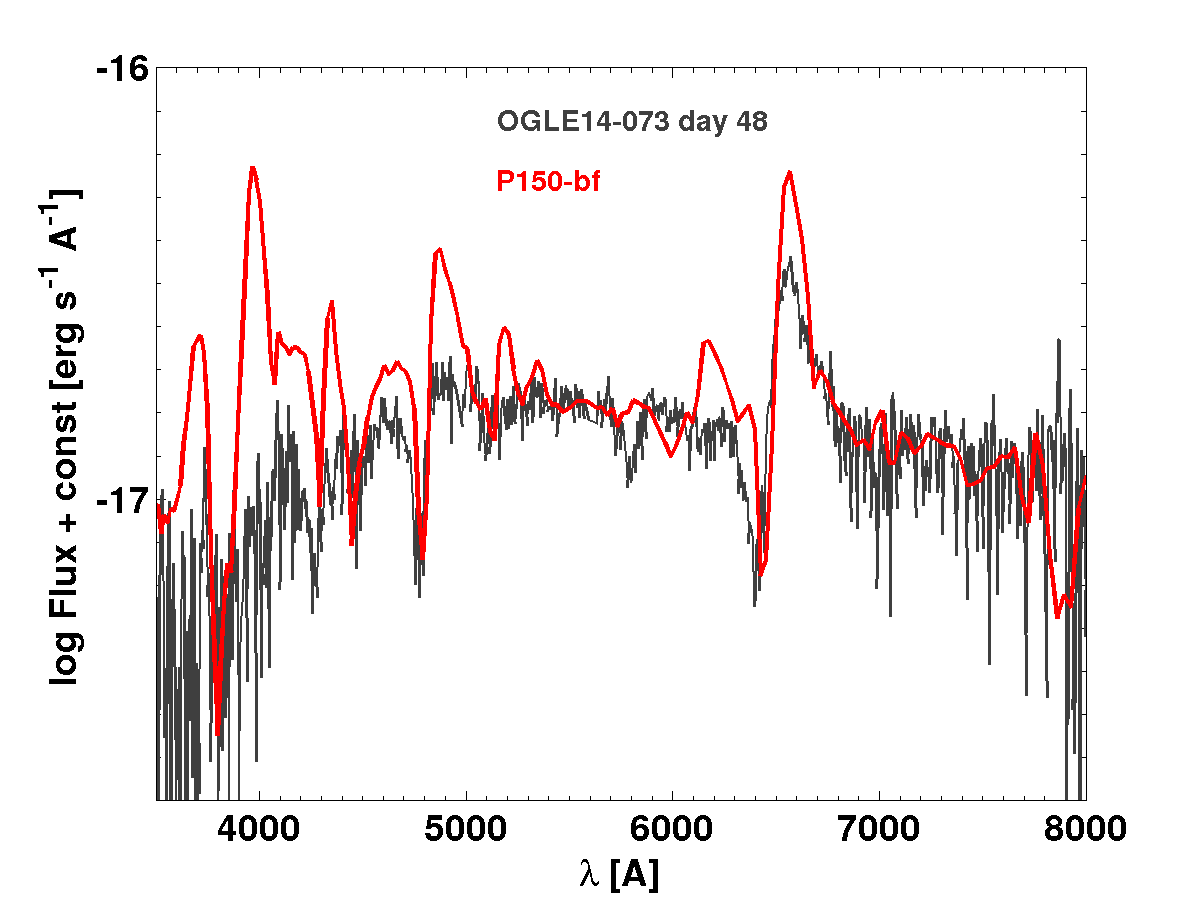}~
\includegraphics[width=\columnwidth]{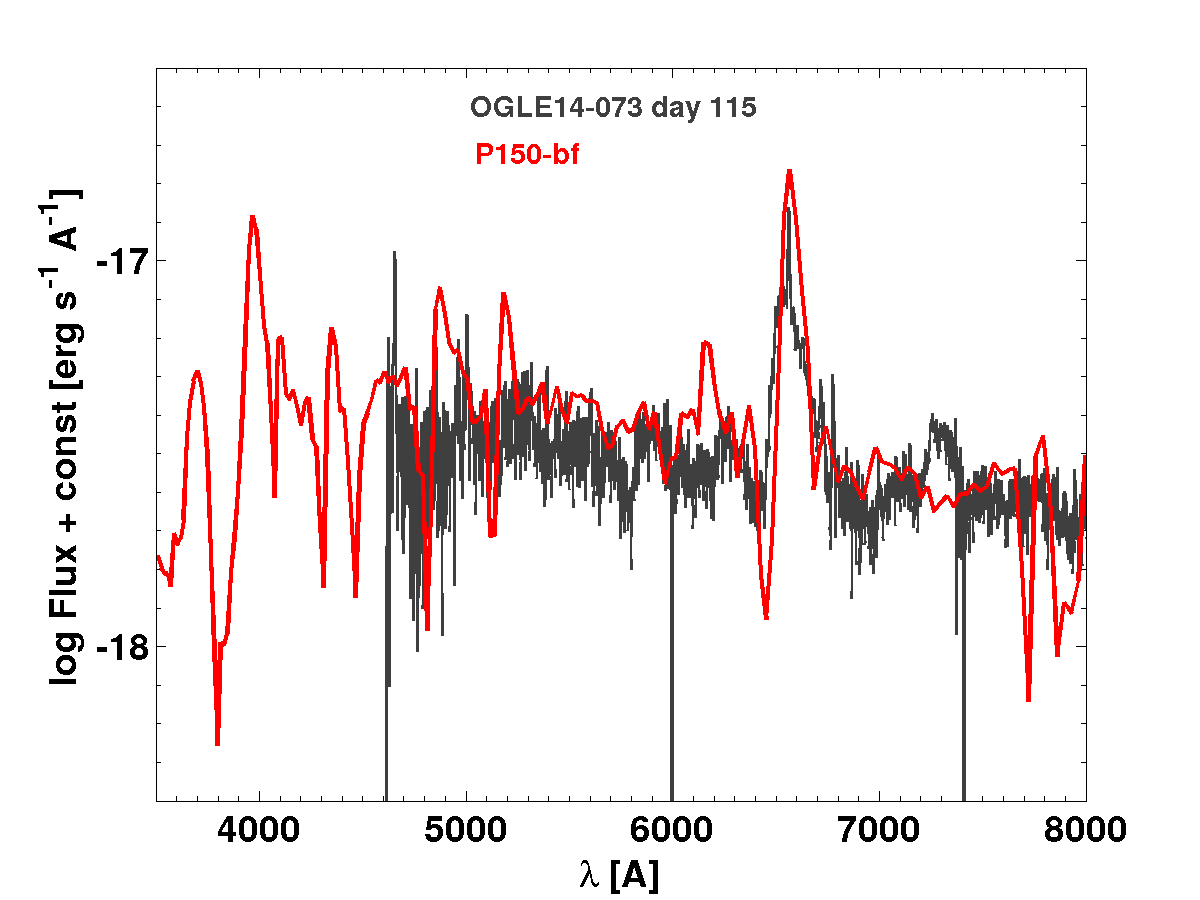}
\caption{Synthetic spectra for model P150-bf (red) at
    day~--26, --7, 48, and 115 relative to peak epoch. For
    comparison the observed spectra of OGLE14-073 at corresponding epochs are
    shown in black. The strongest features in the simulation are indicated in
  the day -7 spectrum.}
\label{figure:spectra}
\end{figure*}

\begin{figure*}
\centering
\includegraphics[width=\columnwidth]{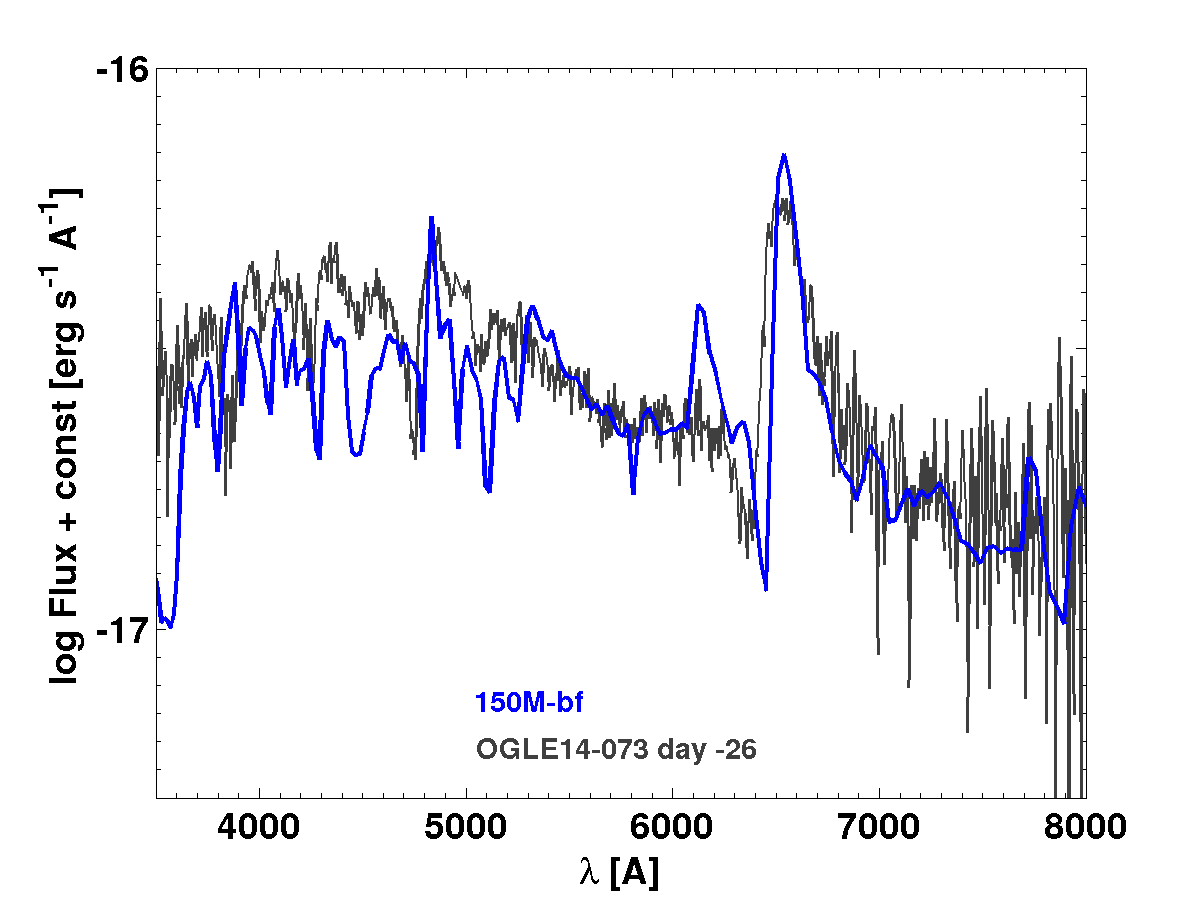}~
\includegraphics[width=\columnwidth]{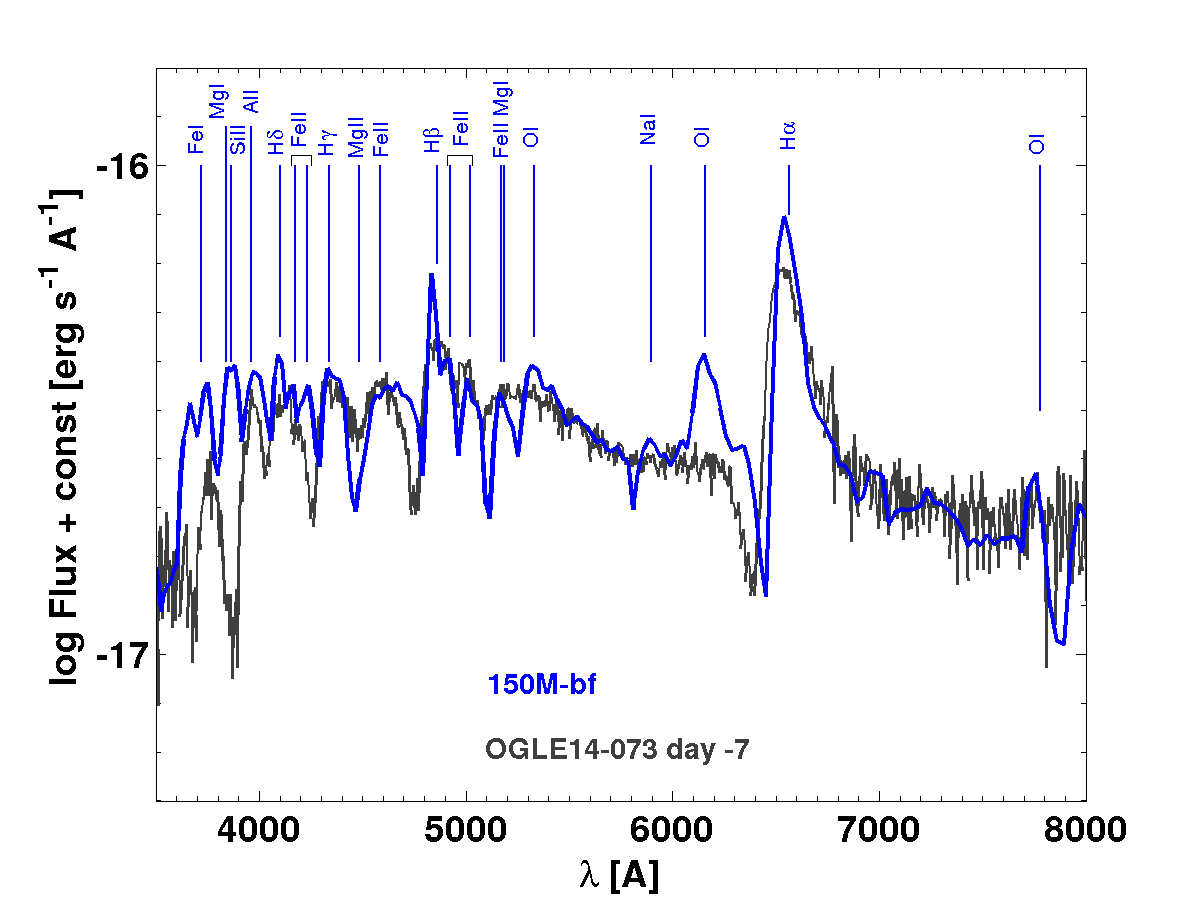}\\
\includegraphics[width=\columnwidth]{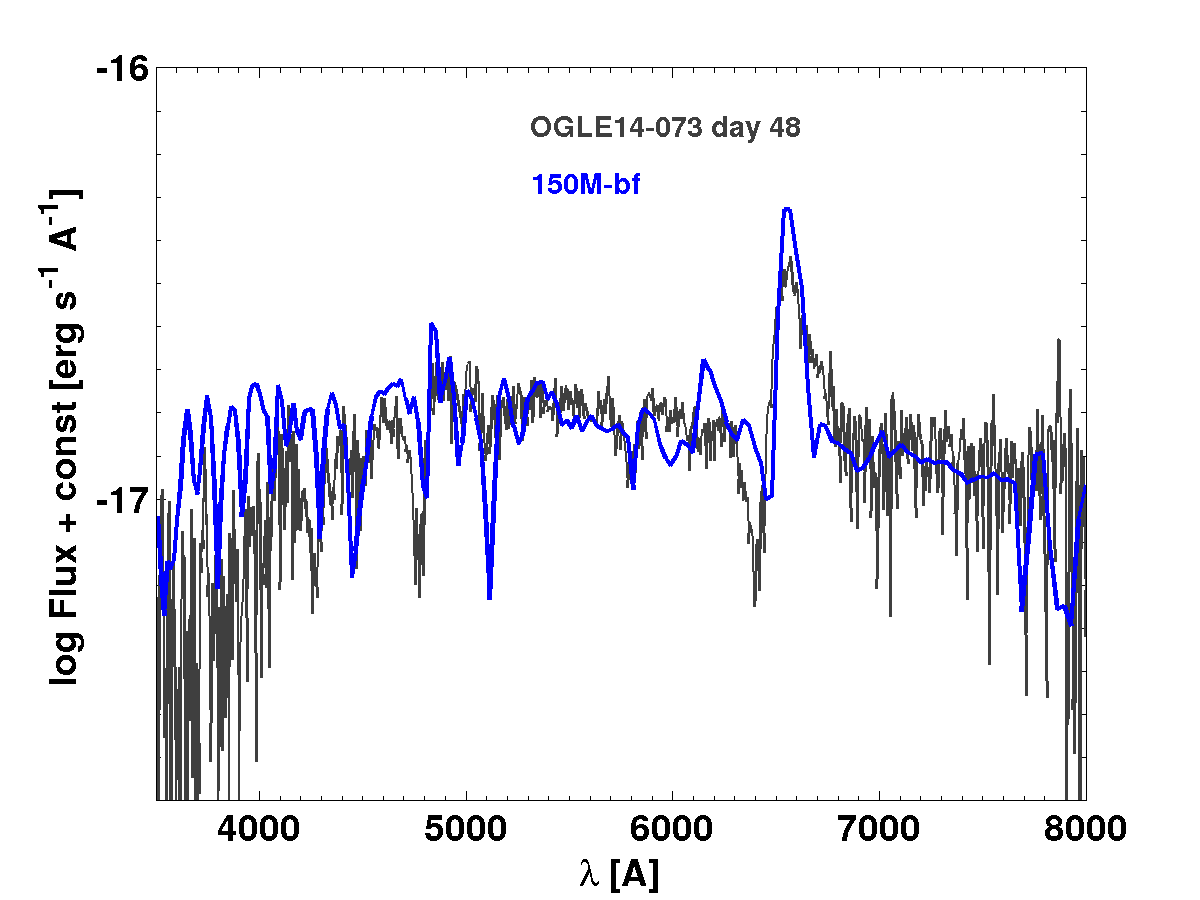}~
\includegraphics[width=\columnwidth]{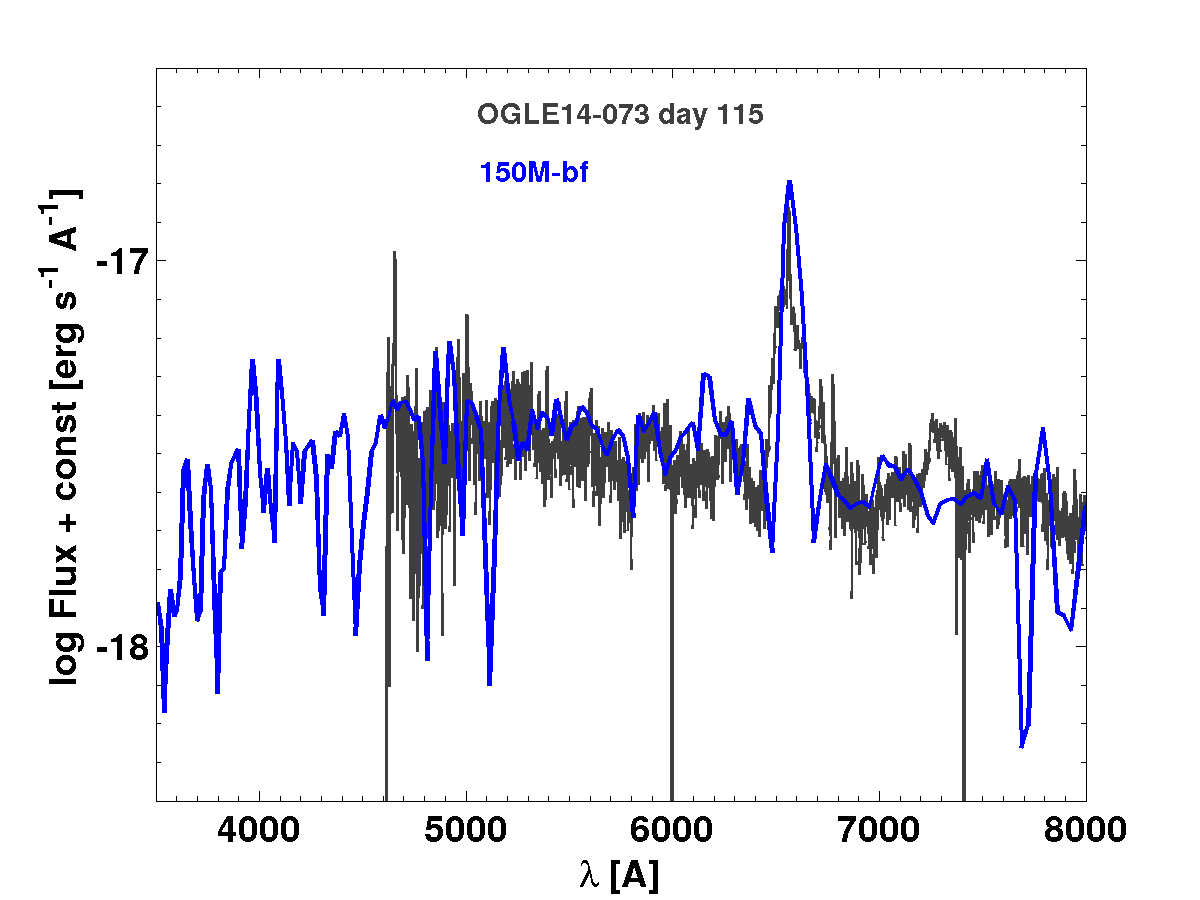}
\caption{Same as Figure~\ref{figure:spectra} but for model 150M-bf,
which is shown in blue.}
\label{figure:spectra_150M-bf}
\end{figure*}

For our best-fit models P150-bf and 150M-bf we simulated the
  spectral evolution from 50 to 450\,d with \textsc{artis} and
  compared the synthetic spectra to those observed for OGLE14-073. A
  subset of these comparisons for both models is shown in
  Figures~\ref{figure:spectra} and \ref{figure:spectra_150M-bf},
  respectively.

Figure~\ref{figure:spectra} shows synthetic spectra of model
  P150-bf for four epochs at day --26, --7, 48, 115 relative to bolometric 
  peak (the corresponding model epochs
  are 189, 208, 263, and 330 days past explosion, respectively).
  Overall the agreement between P150-bf and OGLE14-073 is fairly
  good. The shape of the continuum in the model matches the
observation between 4000\,\AA{} and 8000\,\AA{}. However, there are
minor (0.1~dex) differences in the flux level which corresponds to
an uncertainty in the distance estimate of about 15\,\%{}.  The characteristic
hydrogen lines are clearly visible in the synthetic spectra, and the
shape of H$\alpha$, H$\beta$, and H$\gamma$ are in rough agreement
with the shape of the observed line features in OGLE14-073. A closer
inspection reveals that the detailed shape of the synthetic H
lines is slightly too narrow compared to the observations. This may
suggest insufficient ejecta velocities and in turn too low kinetic
energies in the explosion. However, hydrogen at higher velocity, i.e.\
at a larger mass coordinate, leads to a different light curve
evolution inconsistent with the observed light curve of OGLE14-073.
Uncertainties in the ionization/excitation structure may also affect
the width of the hydrogen lines.

A clear difference between P150-bf and OGLE14-073 is visible
to the blue of H$\alpha$. While the model spectrum shows an emission
feature at around 6200\,\AA\ that is originating from O\,{\sc i}
$\lambda6159$, no such feature is present in OGLE14-073 at --26, --7
and 48 days. A similar behavior is observed for O\,{\sc i}
$\lambda5332$ that leads to a clear absorption feature in the model at
around 5200\,\AA\, but not in OGLE14-073. In contrast, both model and
data show an absorption feature at around 5100\,\AA{}. In the model
this feature is due to Mg\,{\sc i} and its strength increases with
time, while in the data it is getting weaker.

Another difference between P150-bf and the observed spectra is
visible in the Ca H\&K absorption. Although present in OGLE14-073, the
features are much weaker than those predicted from our model. This
suggests too much Ca in the model.  The major fraction of calcium is
produced in the pair-instability explosion due to explosive oxygen and
silicon burning, resulting in 0.4~\Msun{} of Ca.  However, there
is also a small contribution resulting from progenitor
metallicity.
This leads to a Ca mass fraction at the $10^{\,-6}$ ($1.5\times10^{\,-6}$)
level up to the outermost ejecta layers (c.f.\
Figure~\ref{figure:chemie}), which owing to the intrinsic strength of the Ca
H\&K features gives rise to significant absorption.  Also, there is a strong Ca
infrared triplet feature in all synthetic spectra. In the corresponding
spectral range, only in the last epoch data is available and again does not
show such a strong feature.

Figure~\ref{figure:spectra_150M-bf} shows synthetic spectra of
  model 150M-bf for four epochs at day --26, --7, 48, 115 relative to
  bolometric peak  (the
  corresponding model epochs are 179, 198, 253, and 320 days past
  explosion, respectively). Similar to P150-bf, model 150M-bf
  also provides a reasonable match with OGLE14-073. The overall flux
  distribution in the the observed range between 4000\,\AA{} and
  8000\,\AA{} agrees fairly well, and the H Balmer lines, which are
  the dominant line features in OGLE14-073, are clearly present in the
  model. As for P150-bf, the detailed line shapes of the Balmer lines
  of 150M-bf are also too narrow compared to OGLE14-073. In fact
  150M-bf has even narrower lines than P150-bf since the maximum
  ejecta velocity in 150M-bf is lower (11,000 vs. 16,700
  km\,s$^{-1}$). The strong emission feature at around 6200\,\AA\ that
  is not seen in OGLE14-073 and originates from O\,{\sc i}
  $\lambda6159$ in our models is also present in 150M-bf.

Apart from these similarities between the two models, there
  are also important differences. While P150-bf shows a strong
  absorption in the Ca\,{\sc ii} H\&K lines, this feature is absent in
  150M-bf. This results from a different treatment of the progenitor 
  metallicity in the initial models 150M and P150 (see 
  Section~\ref{subsect:method1}), which leads to a different Ca
  distribution in the two
  models (see~Figure \ref{figure:chemie}). In model 150M-bf the Ca mass 
  fraction drops to zero at a
  mass coordinate of $\sim10.5$\,M$_\odot$. In contrast, P150-bf contains Ca
  throughout the ejecta, though at a low
  mass fraction of $1.5\times10^{\,-6}$ in the outer layers. However,
  given the large ejecta mass and the high intrinsic strength of the
  Ca\,{\sc ii} H\&K lines, this is sufficient to lead to strong Ca
  features in P150-bf.

Another difference between P150-bf and 150M-bf is the
  distribution of iron-group elements, specifically $^{56}$Ni. In
  P150-bf the $^{56}$Ni mass fraction drops abruptly at 1.6 solar
  masses, while 150M-bf shows a more
  gentle decline of $^{56}$Ni (see Figure~\ref{figure:chemie}). This leads to
  significant contributions from Fe lines (from the decay of
  $^{56}$Ni) in the synthetic spectra of 150M-bf, while the optical
  spectra of P150-bf are still dominated by line features of
  intermediate mass elements.

Despite some problems in the features associated with a few elements,
we emphasize that the synthetic spectra of our best-fitting models provide
overall a good match to the observations of OGLE14-073.
In particular, we find good agreement in the continuum and thus also the
broad-band colors. This is different from previously published PISN models
\citep[e.g.][]{2013MNRAS.428.3227D,2015ApJ...799...18C,2016MNRAS.455.3207J},
which, compared to observed SNe, suffered from a lack of flux at UV and
blue wavelengths owing to strong line blanketing.  Altogether, the good
agreement found from our spectral and light curve modeling make OGLE14-073 a
very promising PISN candidate.

\section[Conclusion]{Conclusion}\label{sect:conclusion}

In this work, we have examined a possible PISN origin of the
bright type\,II supernova OGLE14-073.  As a starting point for our
investigation, we use two first-principle PISN models with an initial
progenitor mass 150~\Msun{} at metallicity $Z=$0.001. Modifications in
the progenitor properties and ejecta structure enable a good match of
the characteristic bolometric features of OGLE14-073.  In particular,
a relatively large mass of $^{56}$Ni, in the range of
1.3\,--\,1.4~\Msun{}, was required to match the peak brightness. This
increase is reasonable given the high sensitivity of the amount of
$^{56}$Ni synthesized during a PISN on small changes in the progenitor
CO core mass as discussed in the introduction.  Furthermore, hydrogen
must be mixed into the inner ejecta regions (down to a radial mass
coordinate of 10~\Msun{}) to produce the characteristic dome-like
shape of the OGLE14-073 light curve around peak. Such mixing is
predicted by multidimensional explosion simulations and observed in
SN~1987A, which shares some similarities with OGLE14-073. The
best-fitting models thus constructed produce synthetic light curves
which match the observed bolometric evolution of OGLE14-073 fairly
well. In particular, the dome-like structure around peaks is
reproduced, as is the sharp post-peak drop-off and the settling onto
the light curve tail which seems to follow the characteristic
radioactivity decline.

In addition to the good agreement in the bolometric light curve, our models
P150-bf and 150M-bf also reproduce the observed spectral evolution of
OGLE14-073 remarkably well. In particular, we find a good match to the overall
SED and the broad-band colours from epochs well before peak to several weeks
thereafter. This is different from previous PISN models, which suffer from a
flux deficit in the blue wavelength regime owing to strong line blanketing by
large amounts of iron-group elements in the ejecta. This effect is much less
prominent in our models, since $^{56}$Ni, the dominant iron-group species, is
confined to the central region of the ejecta.  Our synthetic spectra also
reproduce many of the observed absorption features of OGLE14-073, specifically
the prominent hydrogen Balmer lines. In detail, however, there are some
differences in the line shapes. Given that the models have not been
tuned to fit the spectra this is to be expected. Also, a more sophisticated
radiative transfer treatment may lead to better agreement in the absorption
features.

In summary, our best-fitting models produce synthetic observables that
fit many characteristic features observed in OGLE14-073. In fact, the
match between predicted and observed signatures reported here is
significantly better than for previous PISN candidates
\citep{2013MNRAS.428.3227D,2017MNRAS.464.2854K,2017ApJ...835...13J}.
Consequently, we conclude not only that a PISN origin of OGLE14-073 is
possible but that this bright SN is one of the most promising
PISN candidates.  Late-time nebular spectra may substantiate this
interpretation. However, no such data are available for OGLE14-073
which faded fairly quickly after the last observations reported by
\citet{2017NatAs...1..713T}.

\section*{Acknowledgements}

The {\sc stella} simulations were carried out on the DIRAC Complexity
system (grants ST/K000373/1 and ST/M006948/1), operated by the
University of Leicester IT Services, which forms part of the STFC
DiRAC HPC Service (\url{www.dirac.ac.uk}).  For the {\sc artis}
simulations, we gratefully acknowledge the Gauss Centre for
Supercomputing (GCS) for providing computing time through the John von
Neumann Institute for Computing (NIC) on the GCS share of the
supercomputer JUQUEEN \citep{stephan2015a} at J\"ulich Supercomputing
Centre (JSC). GCS is the alliance of the three national supercomputing
centres HLRS (Universit\"at Stuttgart), JSC (Forschungszentrum
J\"ulich), and LRZ (Bayerische Akademie der Wissenschaften), funded by
the German Federal Ministry of Education and Research (BMBF) and the
German State Ministries for Research of Baden-W\"urttemberg (MWK),
Bayern (StMWFK) and Nordrhein-Westfalen (MIWF). MK acknowledges
support from the Klaus Tschira Foundation.  UMN is supported by the
Transregional Collaborative Research Centre TRR 33 ``The Dark
Universe'' of the German Research Foundation (Deutsche
Forschungsgemeinschaft). RH acknowledges support from
EU-FP7-ERC-2012-St Grant~306901, the World Premier International
Research Centre Initiative (WPI Initiative), MEXT, Japan, and the
``ChETEC'' COST Action (CA16117). The authors thank Ryan Wollaeger
(LANL, USA) for additional simulations with the spectral synthesis
code {\sc SuperNu}.




\bibliographystyle{mnras}
\bibliography{references}


\bsp	
\label{lastpage}
\end{document}